\documentclass[12pt]{jpconf}
\usepackage{amsmath}
\usepackage{amssymb}
\usepackage{amsthm}
\usepackage{amstext}
\usepackage{array}
\usepackage{xypic}
\usepackage{bm}
\usepackage[T1]{fontenc}
\usepackage[cp1250]{inputenc}

\eqnobysec

\newcommand{\AB}{\allowbreak}

\newcommand{\ali}[2]{\mathop{\mathfrak{#1}(#2)}\nolimits}

\newcommand{\ADA}[1]{\ifmmode \ad(#1) \else $\ad(#1)$\fi}
\newcommand{\LI}[2]{\ifmmode#2_1,\AB\,\ldots,\,\AB #2_{#1}%
\else$ #2_1,\AB\,\ldots,\,\AB#2_{#1}$\fi}

\newcommand{\su}[1]{\ali{su}{#1}}
\newcommand{\sltwo}{\ifmmode \ali{sl}{2} \else $\ali{sl}{2}$\fi}

\long\def\comment#1{}

\newcommand{\bMA}[1]{\[\begin{array}{#1}}
\newcommand{\eMA}{\end{array}\]}

\newcommand{\C}{{{\mathbb C}}}

\newcommand{\I}{\mathbb{I}}

\def\be{\begin{equation}}
\def\ee{\end{equation}}

\def\cpn{{\C P^{N-1}}}
\def\bp{{\bar{\partial}}}
\def\p{{\partial}}

\def\tr{{\mathrm{tr}}}

\newcommand{\bea}{\begin{eqnarray}}
\newcommand{\eea}{\end{eqnarray}}
\newcommand{\la}{\lambda}
\newcommand{\ams}{AMS classification scheme numbers: }
\renewcommand{\c}{\cdot}

\begin{document}

\title{On the surfaces associated with $\cpn$ models}

\author{P. P. Goldstein$^1$ and A. M. Grundland$^2$}

\address{$^1$ Theoretical Physics Department, \\
The Andrzej Soltan Institute for Nuclear Studies, \\
Hoza 69, 00-681 Warsaw, Poland}
\ead{Piotr.Goldstein@fuw.edu.pl}
\vspace{5mm}
\address{$^2$ Centre de Recherches Math{\'e}matiques, Universit{\'e} de Montr{\'e}al, \\
C. P. 6128, Succ.\ Centre-ville, Montr{\'e}al, (QC) H3C 3J7, Canada \\
Universit\'{e} du Qu\'{e}bec, Trois-Rivi\`{e}res CP500 (QC) G9A
5H7, Canada} \ead{grundlan@crm.umontreal.ca}


\begin{abstract}
We study certain new properties of 2D surfaces associated with the
$\cpn$ models and the wave functions of the corresponding linear
spectral problem. We show that $\su{N}$-valued immersion functions
expressed in terms of rank-1 orthogonal projectors are linearly
dependent, but they span an $(N\! -\! 1)$-dimensional subspace of
the Lie algebra $\su{N}$. Their minimal polynomials are cubic,
except for the holomorphic and antiholomorphic solutions, for
which they reduce to quadratic trinomials. We also derive the
counterparts of these relations for the wave functions of the
linear spectral problems. In particular, we provide a relation
between the wave functions, which results from the partition of
unity into the projectors. Finally, we show that the angle between
any two position vectors of the immersion functions, corresponding
to the same values of the independent variables, does not depend
on those variables.
\end{abstract} \ams{53A07, 53B50, 53C43, 81T45}


\section{Introduction}
\label{sec:Intro} Over the past decades there has been a
significant progress in the study of immersion of 2D surfaces in
multidimensional Euclidean spaces obtained from $\cpn$ models. The
most fruitful approach to this subject has been achieved through
the description of these surfaces in terms of the homogeneous
variables $f_k$ \cite{Din,EW,Sa,WZ} and orthogonal projectors
$P_k$ \cite{MikZ,GU2,GY1,GG1,ZH}. Using this language, we have
established the recurrence relations for the projectors satisfying
the $\cpn$ model equations, for the wave functions of their
spectral problems and consequently the immersion functions of 2D
surfaces in the Lie algebra $\su{N}$. In this paper we add certain
new properties, concerning both the immersion functions and the
wave functions, in order to enhance the algebraic and geometric
characterization of the studied surfaces. \noindent The surfaces
are defined by a contour integral \cite{GY1}
\be\label{X}
X_k(\xi,\bar{\xi})=i\int_{\gamma}\left(-[\p P_k,P_k]d\xi+[\bp
P_k,P_k]d\bar{\xi}\right),\quad k=0,...,N\! -\! 1,
\ee
which is independent of the path of integration $\gamma\in
\mathbb{C}$ according to the dynamics of the orthogonal rank-1
projectors $P_k$. The projectors $P_k,~0\le k\le N\! -\! 1$, are
successive solutions \cite{WZ} of the Euler-Lagrange (E-L)
equations in the form of a conservation law \cite{MikZ}
\be\label{cons-law}
\p\,[\bp P,P]+\bp\,[\p P,P]=\mathbf{0},
\ee
corresponding to the action integral
\be
\label{action2} \int d\xi d\bar{\xi}\,\mathcal{L}=\tr(\p P\cdot\bp
P)
\ee
with the constraint
\be\label{projective}
P^2=P.
\ee
Equation \eqref{cons-law} ensures that the integrand of \eqref{X}
is an exact differential. This mapping of an area on a Riemann
sphere $S^2$ into a set of $\su{N}$ matrices:
$\Omega\ni(\xi,\bar{\xi})\mapsto X_k(\xi,\bar{\xi})\in
\su{N}\simeq\mathbb{R}^{N^2-1}$ is a generalised Weierstrass
formula for immersion (GWFI) of 2D surfaces in the Euclidean space
$\mathbb{R}^{N^2-1}$ \cite{Kono1,Kono2,NS}. The target spaces of
the projectors $P_k$ are one-dimensional vector functions
$f(\xi,\bar{\xi})\in \mathbb{C}^N$, constituting an orthogonal
basis in $\mathbb{C}^N$ \cite{Din,WZ}
\be
P_k=[1/(f_k^\dagger\!\c\! f_k)]f_k\otimes f_k^{\dagger},
\ee
All the projectors may be obtained from the first projector $P_0$,
whose target space is an arbitrary holomorphic vector function
$f_0(\xi)$, by the recurrence formulae derived in our previous
work \cite{GG1}. The projectors are orthogonal to each other and
they constitute a partition of unity \cite{Din, WZ}
\be\label{orthogonal}
P_k P_l=\delta_{kl}P_k~\quad\text{(no summation),}
\ee
\be\label{partition}
\sum\limits_{k=0}^{N-1}P_k=\I,
\ee
where $\I$ is the $N$-dimensional unit matrix.

For the surfaces corresponding to the projectors $P_k$, the
integration may be performed explicitly, with the result
\cite{GY1}
\be
\label{XfromP}
X_k=-i\left(P_k+2\sum\limits_{j=0}^{k-1}P_j\right)+i
c_k\mathbb{I}, \quad c_k=\frac{1+2k}{N},\quad 0\le k\le N\! -\! 1.
\ee
In this paper we use the inverse formulae for the projectors in
terms of the surfaces $X_k$, obtained in our previous work
\cite{GG1}
\be\label{Pk from Xk}
P_k={X_k}^2-2i\left(c_k-1\right)X_k-c_k
\left(c_k-2\right)\mathbb{I}.
\ee
to turn the projective property, the partition of unity and the
orthogonal property into corresponding properties of the surfaces.
Namely, we obtain the minimal polynomials, dimensionality of the
spanned subspace of $\mathbb{R}^{N^2-1}$, and the angle between
the position vectors of the surfaces, respectively.

In a similar manner we obtain the inverse formulae for the
projectors in terms of the wave functions and the spectral
parameter of the spectral problems. These relations allow us to
derive the corresponding relations for the wave functions. In
particular, we determine the minimal polynomial and a
linear-dependence equation for those functions.

\section{Projectors and soliton surfaces}

In order to obtain the relations between surfaces $X_k$, it is
convenient to express them in terms of the orthogonal rank-1
projectors $P_k$. Making use of the expression \eqref{XfromP} and
the partition of unity in terms of the projectors
\eqref{partition}, it can be shown that the algebraic condition
\be
\sum\limits_{k=0}^{N-1}(-1)^k X_k=\mathbf{0} \label{ind}
\ee
holds. This means that the $\su{N}$-valued immersion functions $X_j$ are linearly dependent.\\
The equation (\ref{ind}) follows directly from the GWFI
(\ref{XfromP}) in terms of projectors $P_k$ and the decomposition of
unity (\ref{partition}). Indeed, subtracting \eqref{XfromP} for
neighbouring $k$, we obtain
\be
X_k-X_{k-1}=-i(P_k+P_{k-1})+\frac{2i}{N}\I. \label{im1}
\ee
By adding the equations \eqref{im1} for every second $k$, we
obtain an equation containing the sum of all the projectors $P_k$,
which is the unit matrix, according to \eqref{partition}. The
final result proves to be exactly eq. \eqref{ind}.

Note that we can obtain the projectors $P_k$ from the surfaces
$X_k$ not only as quadratic functions of the surfaces \eqref{Pk
from Xk}, but also as linear combinations of the surfaces
$X_0,\ldots, X_k$ \cite{GG1}
\be
\label{PkX} P_k=i\sum\limits_{j=1}^k
(-1)^{k-j}\left(X_j-X_{j-1}\right)+(-1)^k i
X_0+\frac{1}{N}\mathbb{I}.
\ee
Thus we may regain all the projectors $P_0,...,P_{N-1}$ from the
appropriate linear combinations of the surfaces  $X_0,...,X_{N-1}$
and the unit matrix. This means that these surfaces span an
$(N-1)$ dimensional subspace of the $\su{N}$ Lie algebra, as the
projectors are linearly independent.

The projective property $P_k^2=P_k$ imposes an algebraic
constraint on the surfaces $X_k$. To find the lowest order
constraint on $X_k$, we compare ${P_k}\c X_k$ obtained from
\eqref{Pk from Xk} multiplied $X_k$ with ${P_k}\c X_k$ obtained
from \eqref{XfromP} multiplied by $P_k$. This yields a cubic
matrix equation
\be\label{3-deg-cond}
(X_k-ic_k\I)[X_k-i(c_k-1)\I][X_k-i(c_k-2)\I]=\mathbf{0},\quad
0<k<N\! -\! 1.
\ee
For holomorphic $(k\! =\! 0)$ and antiholomorphic $(k=N\! -\! 1)$
solutions of the $\cpn$ equation (\ref{cons-law}) the minimal
polynomial for the matrix-valued functions $X_k$ is quadratic.
Namely, for the surfaces corresponding to the holomorphic
solutions we have
\be
(X_0-ic_0\I)[X_0-i(c_0-1)\I]=\mathbf{0},\quad k=0 \label{quad}
\ee
and, using $c_0+c_{N-1}=2$, we get
\be
(X_{N-1}+ic_0\I)[X_{N-1}+i(c_0-1)\I]=\mathbf{0},\quad k=N-1
\label{quadr}
\ee
for the antiholomorphic ones. Although equation (\ref{quadr}) is
apparently the Hermitian conjugate of the equation (\ref{quad}),
their solutions do not have to be the Hermitian conjugates of each
other.

Condition \eqref{3-deg-cond}, as well as two other conditions
(\ref{quad}) and (\ref{quadr}), have simple interpretation if we
diagonalize them (which is always possible, as the matrices are
anti-Hermitian). It follows that the following numbers are
eigenvalues of $X_k~~k=1,...,N-2$
\be\label{eigenvalues}
i c_k, \quad i(c_k-1) \quad \text{and}\quad i(c_k-2),
\ee
while only the first two are eigenvalues for $k=0$ and only the last
two for $k=N-1$.

The three values listed in \eqref{eigenvalues} are the only
eigenvalues of $X_k$. More precisely
\begin{itemize}
\item
The non-degenerate eigenvalue $i\,(c_k-1)$  occurs at every
$X_k,~~k=0,...,N-1$; the corresponding eigenvector is $f_k$ (the
latter follows directly from \eqref{XfromP} and from the fact that
$f_l,~l=0,...,N-1$ are eigenvectors of the projectors $P_k$ with the
eigenvalue $\delta_{kl}$).
\item
The $k$-fold degenerate eigenvalue $i\,(c_k-2)$ occurs at every
$X_k$, except for $k=0$; the $k$ corresponding eigenvectors are
$f_0,...,f_{k-1}$.
\item
The $(N-1-k)$-fold degenerate eigenvalue $ic_k$ occurs at every
$X_k$, except for $k=N-1$; the corresponding eigenvectors are
$f_{k+1},...,f_{N-1}$.
\end{itemize}

Equation \eqref{3-deg-cond}, together with (\ref{quad}) and
(\ref{quadr}), constitute the lowest degree constraints on the
immersion functions $X_k$ of the surfaces (direct substitution of
\eqref{Pk from Xk} into the projective property would yield a
$4^\mathrm{th}$ degree one). Although equation \eqref{3-deg-cond}
is obvious when we look at the source of $X_k$ \eqref{XfromP}, it
is nevertheless a nontrivial constraint imposed on the surfaces.
Since all the eigenvalues are independent of the coordinates
$(\xi, \bar{\xi})$, the whole kinematics of a moving frame may
only be due to variation
of the diagonalizing (unitary) matrix.\\

\noindent Let us now present certain geometrical aspects of
surfaces immersed in the $\su{N}$ Lie algebras. Once we have the
immersion functions of the surfaces, we can describe their metric
and curvature properties.
\begin{enumerate}
\item
Let $g_k$ be the metric tensor corresponding to the surface $X_k$.
Its components will be marked with indices outside the parentheses
to distinguish them from the index of the surface. Then the
diagonal elements of the metric tensor are zero. This property
directly follows from vanishing of $\tr(\p P_k \p P_k)$ and its
Hermitian conjugate, proven in \cite{GG1}
\be\label{g11=0}
\begin{split}
&(g_k)_{11}=(\p X_k, \p X_k)=\frac{1}{2}\tr([\p
P_k,P_k]\c [\p P_k,P_k])=-\frac{1}{2}\tr(\p P_k\c\p P_k)=0,\\
&(g_k)_{22}=(\bp X_k, \bp X_k)=\frac{1}{2}\tr([\bp P_k,P_k]\c [\bp
P_k,P_k])=-\frac{1}{2}\tr(\bp P_k\c\bp P_k)=0,
\end{split}
\ee
where the inner product $(A,B)$ of the $\su{N}$ matrices is
defined by \cite{GSZ}
\be\label{scalar}
(A,B)=-\frac{1}{2}\tr(A\c B).
\ee
\item
The nonzero off-diagonal element $(g_k)_{12}=(g_k)_{21}$ is equal
to
\be\label{g12}
(g_k)_{12}=-\frac{1}{2}\tr(\p X_k\c\bp X_k)=-\frac{1}{2}\tr([\p
P_k,P_k]\c [\bp P_k,P_k])=\frac{1}{2}\tr(\p P_k\c\bp P_k).
\ee
Thus the first fundamental form reduces to
\be\label{first}
I_k=\tr(\p P_k\c\bp P_k)\,d\xi d\bar{\xi}.
\ee
The second fundamental form
\be
\label{second} II_k=(\p^2 X_k-(\Gamma_k)^1_{11}\p
X_k)d\xi^2+2\p\bp X_k d\xi d\bar{\xi}+(\bp^2
X_k-(\Gamma_k)^2_{22}\bp X_k)d\bar{\xi}^2,
\ee
is easy to find when we determine the Christoffel symbols
$(\Gamma_k)^1_{11}$ and $(\Gamma_k)^2_{22}$. These are the only
nonzero components of $\Gamma_k$. From equation \eqref{g12}, we
get
\be\label{Chris}
(\Gamma_k)^1_{11}=\p\ln{(g_k)_{12}},\quad
(\Gamma_k)^2_{22}=\bp\ln{(g_k)_{12}}.
\ee
Using \eqref{X} and the E-L equations \eqref{cons-law} together with
\eqref{Chris}, we can write \eqref{second} as
\be
\begin{split}
II_k&= -\tr(\p P_k\c\bp P_k)\,\p\frac{[\p P_k,P_k]}{\tr(\p
P_k\c\bp P_k)}d\xi^2+2i\,[\bp P_k,\p P_k]d\xi d\bar{\xi}\\&+\tr(\p
P_k\c\bp P_k)\,\bp\frac{[\bp P_k,P_k]}{\tr(\p P_k\c\bp
P_k)}d\bar{\xi}^2.
\end{split}
\ee
Implementation of the above result for the metric of the surfaces
induced by Veronese solutions of the E-L equations
\eqref{cons-law} is presented in detail in \cite{GG1}.
\end{enumerate}

Also the following $2^\mathrm{nd}$ order differential conditions
hold:
\begin{eqnarray}
(\p\bp X_k,\p X_k)=0,& & \qquad (\p\bp X_k,\bp X_k)=0\label{cd1},\\
(\p\bp X_k,\p^2 X_k)=0,& & \qquad (\p\bp X_k,\bp^2
X_k)=0\label{cd2}.
\end{eqnarray}
Equations \eqref{cd1} follow from direct differentiation of
\eqref{g11=0}. Hence, the mixed derivatives of the matrices $X_k$
coincide and are normal to the surfaces \cite{GSZ}. The second
order differential constraints (\ref{cd2}) are calculated
straightforwardly from the definition
\be
\begin{split}
&(\p\bp X_k,\p^2 X_k)=-\frac{1}{2}\tr([\bp P_k,\p P_k]\cdot [\p^2
P_k,P_k])=-\frac{1}{2}\big{[}\tr(\bp P_k\p P_k\p^2
P_kP_k)\\&-\tr(\p P_k\bp P_k\p^2P_kP_k)-\tr(\bp P_k\p P_k P_k\p^2
P_k)+\tr(\p P_k\bp P_k P_k\p^2 P_k)\big{]}=0,
\end{split}
\ee
since the conditions
\be
P_k\bp P_k \p P_k=\bp P_k \p P_k P_k,\qquad P_k\p P_k\bp P_k=\p P_k\bp P_k P_k,
\ee
hold. Similarly, the second relation in (\ref{cd2}) holds for its
respective Hermitian conjugates. Note that equations (\ref{cd1})
and (\ref{cd2}) are gauge-invariant since they are expressed in
terms of the projectors $P_k$.

We now show that the surfaces $X_k$, $X_l$ do not have common
points for $k\ne l$, with the exception of $X_0$ and $X_1$ in the
$\C P^1$ model, where simply $X_0$ coincides with $X_1$.

Indeed, let $l>k$ be two different indices of the surfaces.
Subtracting \eqref{XfromP} from the analogous expression for
$X_l$, we obtain
\be\label{differenceX}
-i\,\left[P_l -
P_k+2\sum\limits_{j=k}^{l-1}P_j-\frac{2(l-k)}{N}\I\right]=\mathbf{0}
\ee
Multiplying eq. \eqref{differenceX} by $P_k$, we obtain
\be\label{dif-k}
P_k\left[1-\frac{2(l-k)}{N}\right]=\mathbf{0}.
\ee
On the other hand, when we multiply both hand sides of
\eqref{differenceX} by $P_{l-1}$, we get
\be\label{dif-l}
P_{l-1}\left[1-\frac{l-k}{N}\right]=\mathbf{0}\quad\text{for
$k<l-1$, and}
\ee
\be\label{dif-l-1}
P_k\left(1-\frac{2}{N}\right)=\mathbf{0}\quad\text{for $k=l-1$.}
\ee
All of the equations (\ref{dif-k}), (\ref{dif-l}) and
(\ref{dif-l-1}) may only be satisfied when ${N=2},~{l=1},~{k=0}$.
In that case \eqref{ind} yields immediately $X_1=X_0$.

We now show that the immersion functions $X_k,~X_m$ make a constant
angle in the sense of the Euclidean inner product \eqref{scalar},
i.e. the angle $\Phi_{km}$ between the immersion functions $X_k$ and
$X_m$ does not depend on the particular choice of projector $P_0$,
nor on the coordinates $\xi$, $\bar{\xi}$. Namely, the angle
$\Phi_{km}$ between two different functions $X_k$, $X_m$, $k<m$, is
given by the equation
\be
\cos{\Phi_{km}}=\frac{c_k(2-c_m)}{\left\{\left[c_k(2-c_k)-1/N\right]\left[c_m(2-c_m)-1/N\right]\right\}^{1/2}}
\label{cos}
\ee
The formula \eqref{cos} may be obtained in a straightforward way
either by calculating the appropriate scalar product from the GWFI
for $X_k$, $X_m$ \eqref{XfromP} (bearing in mind that the projectors
$P_0$, $P_1$, $\ldots$, $P_k$ are mutually orthogonal), or by direct
operations on the eigenvalues of the immersion functions. In either
way we obtain for $m>k$
\be
(X_k,X_m)=-(1/2)\,\tr(X_k\cdot X_m)=(N/2)\,c_k(2-c_m),
\ee
while
\be
(X_k,X_k)=-(1/2)\,\tr(X_k\cdot X_k)=(1/2)\,[N\,c_k(2-c_k)-1].
\ee
It may easily be proven that \eqref{cos} always yields
$\cos\Phi_{km}\in (0,1)$ unless $N=2$ (and obviously $k=0,~m=1$), for
which the surfaces coincide (and the cosine is obviously equal to 1).
Equation \eqref{cos} is symmetric with respect to a transformation
$k\longleftrightarrow N-1-m$, which may be seen e.g. in the table of
$\cos\Phi_{km}$ for the $\mathbb{C}P^3$ model\\[3mm]
\begin{tabular}{c|cccc}
$~~~k\,\backslash\, m$&$1$&$2$&$3$\\
\hline\noalign{\smallskip}
$0$&$5/\sqrt{33}$&$\sqrt{3/11}$&$1/3$\\
1&&$9/11$&$\sqrt{3/11}$\\
2&&&$5/\sqrt{33}$\\
\end{tabular}
\\[3mm]
We can regard the immersion functions $X_k$ as position vectors,
whose ends draw the two-dimensional surfaces in a
$N^2\!-\!1$-dimensional $\su{N}$ algebra. The above result means that
the position vectors make a constant angle with each other,
independent of the variables $\xi,~\bar{\xi}$. Moreover, the angle is
the same for all choices of the $P_0$ solutions of the E-L equations
\eqref{cons-law} within a particular $\cpn$ model.

\section{Projectors and the spectral problem}
The spectral problem is closely related to the immersion functions
of the surfaces. The relation between the wave functions and the
immersion functions is given by the Sym-Tafel (ST) formula
\cite{Sym1,Sym2,Sym3,Taf}. The wave functions are also related to
the immersion functions by their asymptotic properties. No wonder
that the results of the previous section have their counterparts
in corresponding relations between the wave functions.

Similarly to the surfaces, the wave functions of the spectral problem
can also be expressed in terms of the projectors. The spectral
problem found by Zakharov and Mikhailov \cite{MikZ} reads
\be\label{spectral}
\p \Phi_k=\frac{2}{1+\la}[\p P_k,P_k]\Phi_k , \qquad \bp
\Phi_k=\frac{2}{1-\la}[\bp P_k,P_k]\Phi_k, \qquad k=0,1,\ldots,N-1,
\ee
where $\la\in\mathbb{C}$ is the spectral parameter. An explicit
solution of the linear spectral problem (\ref{spectral}) for which
$\Phi_k$ tends to $\I$ as $\lambda \to \infty$ is given by
\cite{DHZ}
 \be\label{PhifromP}
\Phi_k=\mathbb{I}+\frac{4\la}{(1-\la)^2}\sum\limits_{j=0}^{k-1}P_j-\frac{2}{1-\la}P_k,\quad
{\Phi_k}^{-1}=\mathbb{I}-\frac{4\la}{(1+\la)^2}\sum\limits_{j=0}^{k-1}P_j-\frac{2}{1+\la}P_k,
\ee
where $\la$ is purely imaginary and thus $\Phi_k\in SU(N)$. This
in turn yields the projectors $P_k$ in terms of the wave functions
\cite{GG1}
\be\label{PfromPhi}
P_k=(1/4)\left[2(1+\la^2)\mathbb{I}-(1-\la)^2\Phi_k-(1+\la)^2\Phi_k^{-1}\right].
\ee
The projective property of $P_k$ may be represented in terms of
$\Phi_k$ as a factorisable $4^\mathrm{th}$ degree expression with
one double (squared) factor
\be
P_k^2-P_k=(1/16)\Phi_k^{-2}(\mathbb{I}-\Phi_k)\left[(1+\la)^2
-(1-\la)^2\Phi_k\right]\left[(1+\la)-(1-\la)\Phi_k\right]^2=\mathbf{0}.
\ee
Hence, the minimal polynomials of the matrices $\Phi_k$ are cubic
and they satisfy the equation, resembling the corresponding
equation for the surfaces $X_k$ \eqref{3-deg-cond}, but explicitly
depending on the spectral parameter
\be
(\I-\Phi_k)\left[(1+\lambda)\I-(1-\lambda)\Phi_k\right]\left[(1+\lambda)^2\I-(1-\lambda)^2\Phi_k\right]=\mathbf{0}
\label{cubic1}
\ee
for $1<k<N-1$. Similarly to the surfaces \eqref{quad},
\eqref{quadr}, quadratic matrix equations are sufficient for $k=0$
and $k=N-1$: the equation with only the first two factors of
\eqref{cubic1} is satisfied by $\Phi_0$ and that with only the
last two factors of \eqref{cubic1} by $\Phi_{N-1}$.

The immersion functions $X_k$ may be expressed in terms of the
wave functions $\Phi_k$ in two ways: either by the ST formula
\cite{Sym1,Sym2,Sym3,Taf}
\be\label{ST}
X_k^{ST}=-\frac{i}{2}(1-\lambda^2)\Phi_k^{-1}\p_{\lambda}\Phi_k
\ee
or as a limit \cite{GG1}
\be
X_k=i\lim_{\lambda\to\infty}
\left[\frac{\lambda}{2}\Phi_k+\left(c_k-\frac{\lambda}{2}\right)\I\right].
\label{lim}
\ee
Using equation (\ref{lim}), one can check by explicit calculation
that the cubic minimal polynomial (\ref{cubic1}) for the wave
function $\Phi_k$ coincides with the cubic polynomial
(\ref{3-deg-cond}) for the immersion function $X_k$ in the limit
$\lambda\to\infty$.

We now show that the partition of unity (\ref{partition}) for the
projectors $P_k$ imposes constraints on the wave functions
$\Phi_k$, given by the relation
\be
\sum\limits_{j=0}^{N-1}\Phi_j\left(\frac{1-\lambda}{1+\lambda}\right)^j=\frac{1+\lambda}{2\lambda}\left[1-\left(\frac{1-\lambda}{1+\lambda}\right)^{N-2}\right]\I.
\ee
Indeed, using the wave functions $\Phi_k$, which can be expressed in
terms of the projectors $P_k$ through the formula (\ref{PhifromP}),
for the indices $k$ and $k-1$, we obtain
\be
\sum\limits_{k=0}^{N-1}\sum\limits_{j=0}^{k}(\Phi_{j-1}-\Phi_j)\left(\frac{1+\lambda}{1-\lambda}\right)^{k-j}=\frac{2}{1-\lambda}\I,
\ee
or equivalently
\be
\sum\limits_{j=0}^{N-1}(\Phi_{j-1}-\Phi_j)\left(\frac{1+\lambda}{1-\lambda}\right)^{N-j}-\sum\limits_{j=0}^{N-1}(\Phi_{j-1}-\Phi_j)=\frac{4\lambda}{1-\lambda^2}\I,
\ee
or, factoring out the coefficients with respect to $\Phi_{j-1}$ and $\Phi_j$, we get
\be
\sum\limits_{j=0}^{N-1}\Phi_{j-1}\left(\frac{1+\lambda}{1-\lambda}\right)^{N-j}-\sum\limits_{j=0}^{N-1}\Phi_j\left(\frac{1+\lambda}{1-\lambda}\right)^{N-j}-\Phi_{N-1}=\left(\frac{1+\lambda}{1-\lambda}\right)^2\I.
\ee
Finally, this expression can be written in the form
\be\label{sumPhi}
\sum\limits_{j=0}^{N-1}\Phi_j\left(\frac{1-\lambda}{1+\lambda}\right)^j=
\frac{1+\lambda}{2\lambda}\left[1-\left(\frac{1-\lambda}{1+\lambda}\right)^{N-2}\right]\I.
\ee
Equation \eqref{sumPhi}, transformed into an appropriate equation
for the variable $(\lambda/2)\Phi_j-(c_j-\lambda/2)\I$, turns into
equation \eqref{lim} in the limit $\lambda\to \infty$.
\\
Note that in view of the linear dependence of the immersion
functions $X_k$, i.e. equation (\ref{ind}), the ST formula
\eqref{ST} leads to the following differential constraint on the
wave functions $\Phi_k$
\be
\sum\limits_{j=0}^{N-1}(-1)^jX_j^{ST}=-\frac{i}{2}(1-\lambda^2)\p_{\lambda}\ln\prod_{j=0}^{N-1}\Phi_j^{(-1)^j}=\mathbf{0}.
\ee
This implies that the expression
$\prod_{2l<N}\Phi_{2l}\prod_{2l+1<N}\Phi_{2l+1}^{-1}$ is
independent of the spectral parameter $\lambda$ but it may depend
on the variables $\xi$ and $\bar{\xi}\,\in\mathbb{C}$.

\section{Concluding remarks}
In our work we develop the approach proposed in \cite{GG1}, which
relies on construction of the consecutive surfaces and immersion
functions in terms of projectors. The inverse formulae found in
that work allowed for deriving additional properties of the
surfaces immersed in the $\su{N}$ Lie algebra and functions
immersed in the $SU(N)$ Lie group. In particular
\begin{itemize}
\item
We have shown that the number of linearly independent surfaces,
associated with the $\cpn$ models is $N-1$.
\item
The angles between the position vectors of any two surfaces are
constant (independent of $\xi,~\bar{\xi}$ and independent of the
choice of the holomorphic solution). The surfaces do not intersect
with each other for $\cpn ,~N\ge 2$; the only two surfaces of
$\mathbb{C}P^1$ coincide.
\item
All the surfaces associated with the $\cpn$ models satisfy a
$3^\mathrm{rd}$-degree matrix equation, which reduces to a
$2^\mathrm{nd}$-degree equation for the holomorphic and
antiholomorphic solutions of the E-L equations \eqref{cons-law}.
\item
The corresponding relations hold for the wave functions of the
spectral problem. Moreover the asymptotic properties of those
functions while the spectral parameter tends to infinity connect
the relations for the wave functions with those for the immersion
functions.
\end{itemize}
The proposed approach opens a field of further research for other
sigma models.

\ack 
A.M.G.'s work was supported by a research grant from NSERC of
Canada. This project was completed during A.M.G.'s visit to the
\'Ecole Normale Superieure de Cachan, and he would like to thank
the CMLA for their kind invitation and hospitality.

\section*{References}

\begin{footnotesize}

\end{footnotesize}

\end{document}